\documentclass[doublecol]{epl2}
\usepackage{amsmath,amssymb}
\usepackage{graphicx,soul}

\newcommand{\dd}{\mathrm{d}}

\newcommand{\pd}[2]{\frac{\partial #1}{\partial #2}}
\newcommand{\mean}[1]{\langle #1 \rangle}
\newcommand{\Int}[1]{\int\dd #1\;}
\newcommand{\IInt}[3]{\int_{#2}^{#3}\dd #1\;}

\renewcommand{\vec}[1]{\mathbf #1}
\DeclareMathOperator{\tr}{tr}

\newcommand{\gam}{\gamma}

\newcommand{\kap}{\kappa}
\newcommand{\lam}{\lambda}

\newcommand{\sig}{\sigma}
\newcommand{\Om}{\Omega}
\newcommand{\ra}{\rightarrow}
\newcommand{\rr}{^{(R)}}

\newcommand{\id}{\mathbf 1}
\newcommand{\im}{\text i}
\newcommand{\x}{\vec r}
\newcommand{\kT}{k_\text{B}T}
\newcommand{\z}{\boldsymbol\omega}
\newcommand{\strain}{\boldsymbol\epsilon}
\newcommand{\stress}{\boldsymbol\sig}
\newcommand{\nois}{\boldsymbol\xi}
\newcommand{\sa}{^{(\text{a})}}
\newcommand{\si}{^{(\text{i})}}
\newcommand{\Dr}{D_\text{r}}
\newcommand{\lp}{\ell_\text{p}}

\begin{document}

\title{Stochastic thermodynamics for active matter}

\author{Thomas Speck}
\institute{Institut f\"ur Physik, Johannes Gutenberg-Universit\"at Mainz,
  Staudingerweg 7-9, 55128 Mainz, Germany}

\abstract{The theoretical understanding of active matter, which is driven out
  of equilibrium by directed motion, is still fragmental and model
  oriented. Stochastic thermodynamics, on the other hand, is a comprehensive
  theoretical framework for driven systems that allows to define fluctuating
  work and heat. We apply these definitions to active matter, assuming that
  dissipation can be modelled by effective non-conservative forces. We show
  that, through the work, conjugate extensive and intensive observables can be
  defined even in non-equilibrium steady states lacking a free energy. As an
  illustration, we derive the expressions for the pressure and interfacial
  tension of active Brownian particles. The latter becomes negative despite
  the observed stable phase separation. We discuss this apparent
  contradiction, highlighting the role of fluctuations, and we offer a
  tentative explanation.}


\maketitle


\section{Introduction}

Active matter is a somewhat vague term applied to systems driven out of
thermal equilibrium by converting stored or locally supplied free energy into
directed motion~\cite{vics12}. This notion encompasses systems as disparate as
bacterial suspensions~\cite{wens12} and microtubules connected by motor
proteins~\cite{sanc12}. Of particular interest is the phoretic motion of
colloidal Janus particles~\cite{theu12,pala13,butt13,bial14}. The question
whether active matter could be described in terms of equilibrium
thermodynamical concepts~\cite{taka15} has recently attracted considerable
interest, in particular the definition of a
pressure~\cite{taka14,solo15,solo15a,wink15}. While the conventional approach
to many non-equilibrium problems is to explicitly formulate and
(approximately) solve detailed equations of motion, the power of a
thermodynamic description is that it is independent of many of those details
and predicts universal bounds.

Over the last two decades, stochastic thermodynamics has evolved into a
comprehensive theoretical framework to treat driven systems in contact with a
heat reservoir~\cite{seif12}. It applies chiefly to systems with a clear
separation of degrees of freedom, where entropy production is due to the slow
and observable degrees of freedom following (effective) Markovian stochastic
dynamics. So far it has been applied mainly to study colloidal systems and
biological matter~\cite{coll05,blic06}. The central tenet of stochastic
thermodynamics is the first law, not only for mean values but on the level of
single trajectories~\cite{seki98}. Fluctuations of work and heat are
constrained by fluctuation theorems~\cite{kurc98,lebo99,seif05a}, which
ultimately are a consequence of time-reversal symmetry.

\begin{figure}[t]
  \centering
  \includegraphics{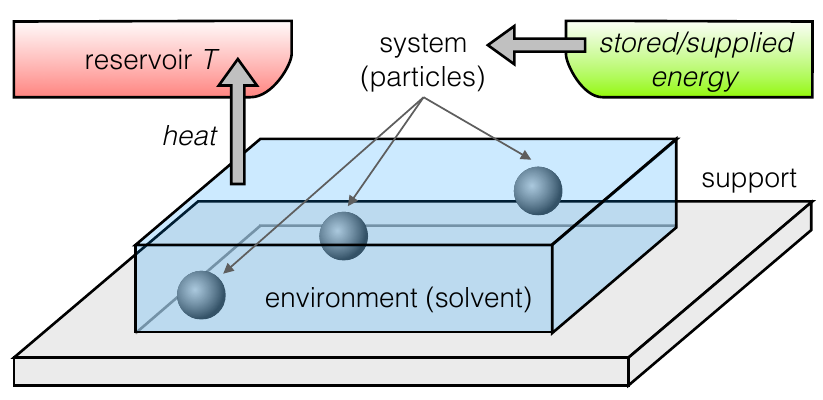}
  \caption{Typical experimental situation. A microfluidic device or a cell
    containing a suspension is mounted on a massive support. We identify the
    suspended ``particles'' (bacteria, colloidal particles, etc.) as the
    system and the solvent (plus the surrounding laboratory) as the
    environment with temperature $T$. The system is externally supplied with
    energy, which is eventually dissipated as heat.}
  \label{fig:sketch}
\end{figure}

In this paper, we aim to describe particles (cells, bacteria, \emph{etc.})
that are moving in an aqueous environment. The particles constitute the
\emph{system} specified by their positions (and possibly internal degrees of
freedom), while the surrounding solvent takes the role of the \emph{reservoir}
with well-defined temperature $T$, see Fig.~\ref{fig:sketch}. The physical
picture is that the reservoir is sufficiently large so that the fluctuations
remain those of an equilibrium solvent. Moreover, we assume that both the
particles and the solvent are enclosed within a container that can take up
arbitrary amounts of momentum without being set in motion. This assumption
certainly holds for colloidal particles and bacteria moving in a chamber
mounted in the laboratory, and indeed for most composite systems attached to
the earth (\emph{e.g.}, bacteria in a pond).

Previous discussions on a possible relation to thermodynamics have focused on
an effective free energy for active particles~\cite{taka15,cate15} and
fluctuation relations~\cite{gang13}. Here we follow a different strategy that
is based on the (virtual) work required to deform a volume of active
matter. The first step is to introduce conjugate observables based on the work
instead of a free energy. We introduce the concept of effective
non-conservative forces for active matter in order to model
dissipation. Finally, we investigate in more detail an inhomogeneous system
due to phase-separation into a dense and dilute phase~\cite{butt13}. Recently,
for this situation it has been found that the interfacial tension (the work to
create the interface) is negative~\cite{bial15} but phase separation is
stable, which can be rationalized within the theoretical framework developed
here.


\section{Stochastic thermodynamics}

Let $\z$ be the configuration of the system with potential energy
$U(\z,\vec X)$, which also depends on a number of external parameters
$\vec X=(X_1,\dots)$ such as volume $V$. These can be controlled externally
while the configuration evolves stochastically due to the coupling with the
solvent. Here we assume a time-scale separation such that momenta (and
possibly fast degrees of freedom) relax very quickly and thermalize on time
scales much shorter than experimentally accessible. The effective Hamiltonian
after integrating out the equilibrated degrees of freedom reads
\begin{equation}
  H(\z,\vec X) = F_\text{id}(T,\vec X) + U(\z,\vec X),
\end{equation}
where $F_\text{id}(T,\vec X)$ is the ideal free energy for non-interacting
particles in the absence of a potential energy.

The work
\begin{equation}
  \label{eq:work}
  \delta\hat w = \pd{H}{\vec X}\cdot\dd\vec X + \vec f\cdot\dd\z
\end{equation}
is composed of two terms: the first term takes into account the work that is
spent due to a change of external parameters whereas the second term arises
due to non-conservative forces $\vec f(\z,\vec X)$. The hat emphasizes that
this expression depends on the microstate, while the $\delta$ emphasizes that
both work and heat are not exact differentials. The heat that is dissipated
then follows from the first law of thermodynamics
$\delta\hat q=\delta\hat w-\dd H$ expressing the conservation of energy. The
sign is convention, here heat that is dissipated into the solvent is
positive. We stress that this balance equation only involves energies and as
such is independent of dynamics, in particular hydrodynamic interactions do
not enter. Only when taking the average does the dynamics of the system enter.

However, before it is instructive to recall conventional thermodynamics. The
systems we are interested in are typically described by a fixed number of
constituents in a given volume $V$, the relevant ensemble for which is the
canonical ensemble with free energy $F(T,\vec X)=\mean{H}-TS$, where $S$ is
the entropy and the brackets $\mean{\cdot}$ denote the average. The reversible
work is given by the differential
\begin{equation}
  \delta w = \dd F = \pd{F}{\vec X}\cdot\dd\vec X = \dd\mean{H} - T\dd S.
\end{equation}
This expression corroborates our identification of stochastic work and heat
since $-T\dd S$ is the conventional heat. The work can also be expressed as
$\delta w=-\vec K\cdot\dd\vec X$ with \emph{conjugate} generalized forces
$K_i=-\partial F/\partial X_i$. In the following we will focus on the
extension of two particular conjugate quantities. First, the pressure $p$
conjugate to the volume,
\begin{equation}
  \label{eq:eq:p}
  \delta w = \dd F = -p\dd V,
\end{equation}
and the interfacial tension $\gam$ describing the reversible work
\begin{equation}
  \label{eq:eq:gam}
  \delta w = \dd F = \gam\dd A
\end{equation}
necessary to change the interface area $A$ in an inhomogeneous system at fixed
volume $V$. The signs agree with our physical intuition, the pressure opposes
the further compression of a system while work has to be spent to extend an
interface.


\subsection{Virtual box change and conjugate stress}

We first consider point particles with microstate $\z\equiv\{\x_k\}$ specified
by the positions of all $N$ particles. We aim to calculate the work required
by moving all particles according to $\x_k\ra\x_k'=\vec h\x_k$, which
corresponds to a deformation of the volume $V$ (the ``box''). The entries
$h_{ij}$ of the matrix $\vec h$ are now part of the external parameters
$\vec X$. We first consider non-conservative forces to be absent so that the
stationary state corresponds to thermal equilibrium. The strain due to this
transformation is $\strain=\frac{1}{2}\left(\vec h^T\vec h-\id\right)$ with
volume change $V\ra V'=V(1+\tr\strain)$ for small strain~\cite{landau7}. From
Eq.~(\ref{eq:work}) the work thus becomes
\begin{equation}
  \label{eq:work:box}
  \delta\hat w = V\sum_{ij}\hat\sig_{ij}\dd h_{ij}
\end{equation}
with volume $V$ before the transformation and stress tensor
\begin{equation}
  \hat\sig_{ij}(\z) = \frac{1}{V}\left.\pd{H}{h_{ij}}\right|_{\vec h=\id}.
\end{equation}
Eq.~(\ref{eq:work:box}) is our first central result relating the stress to the
work instead of the free energy. Taking the average and exploiting that
$\stress=\mean{\hat\stress}$ is a symmetric tensor, one recovers the more
conventional expression $\mean{\delta\hat w}=V\tr(\stress\dd\strain)$ for the
work. Note that the average is with respect to the stationary state before the
shape change. Hence, $\delta\hat w$ corresponds to the infinitesimal work for
a \emph{virtual} box change without actually having to perform the
deformation.

The ideal free energy contributes $-N\kT\delta_{ij}$ to the stress. The second
contribution due to the potential energy reads
\begin{equation}
  \left.\frac{1}{V}\pd{U}{h_{ij}}\right|_{\vec h=\id} = 
  \frac{1}{V}\sum_{k=1}^N\left(\pd{U}{\x_k}\right)_i(\x_k)_j
\end{equation}
employing the chain rule, where the subscripts $i$ and $j$ label the vector
component. For example, consider a uniform change of the box size with
$\vec h=\lam\mathbf 1$. Since $\dd\lam=\dd V/(Vd)$ we obtain the work
$\delta\hat w=-\hat p\dd V$ with
\begin{equation}
  \frac{1}{d}\tr\hat\sig = 
  -\frac{N\kT}{V} + \frac{1}{Vd}\sum_{k=1}^N\pd{U}{\x_k}\cdot\x_k = -\hat p(\z),
\end{equation}
which is the microscopic virial expression for the pressure as
expected~\cite{hansen}. Here, $V$ is the controlled extensive quantity and the
pressure $p=\mean{\hat p}$ is the conjugate intensive quantity.


\subsection{Non-conservative forces}

We now consider an infinitesimal transformation of the box for systems driven
into a non-equilibrium steady state due to non-conservative forces. The total
work $\delta\hat w=\delta\hat w_\text{ex}+\delta\hat w_\text{hk}$ from
Eq.~(\ref{eq:work}) can then be split into the ``excess'' work
$\delta\hat w_\text{ex}$ due to the (explicit or virtual) change of the box
(with the relative particle positions fixed) and the ``housekeeping'' work
\begin{equation}
  \label{eq:work:hk}
  \delta\hat w_\text{hk} = \sum_{k=1}^N \vec f_k\cdot\dd\x_k
\end{equation}
due to the evolution of the positions (at fixed box shape). The work due to a
box change now reads
\begin{equation}
  \delta\hat w_\text{ex} = \sum_{ij} \left[ \left.\pd{H}{h_{ij}}\right|_{\vec h=\id}
    + \sum_{k=1}^N (\vec f_k)_i(\x_k)_j \right]\dd h_{ij}.
\end{equation}
Employing Eq.~(\ref{eq:work:box}) with $\delta\hat w_\text{ex}$ the
microscopic stress tensor thus becomes
\begin{equation}
  \label{eq:stress}
  \hat\stress(\z) = -\frac{N\kT}{V}\mathbf 1 + \frac{1}{V}\sum_{k=1}^N\left(\vec
    f_k+\pd{U}{\x_k}\right)\x_k^T.
\end{equation}
At this point we have defined a conjugate stress via the work, which reduces
to the conventional expression in the absence of non-conservative forces (with
$\delta\hat w_\text{hk}=0$).

Again, consider the pressure $p$ in a non-equilibrium steady state now
following from $\delta w_\text{ex}=-p\dd V$. That this pressure is still an
intensive quantity follows from arguments similar to conventional
thermodynamics. Suppose we have a system that is partitioned into two volumes
$V_1$ and $V_2$ with total volume $V$. The additive excess work for changing
the partition is thus
$\delta w_\text{ex}=-p_1\dd V_1-p_2\dd V_2=-(p_1-p_2)\dd V_1$ due to
conservation of the total volume. In a steady state, mechanical stability
requires that such a change costs work, $\delta w_\text{ex}\geqslant 0$.
Hence, $p_1=p_2$ (since the volume change can take both signs) with
$\delta w_\text{ex}=0$, which shows that the pressure is uniform in both
subsystems.


\subsection{Inhomogeneous systems}

Next, we consider an inhomogeneous system. To this end we generalize the
stress tensor Eq.~(\ref{eq:stress}) and consider a subregion $R$ with volume
$V_R$ and sum only over particles $k\in R$ within that subregion. Taking the
average $\stress(\x)=\mean{\hat\stress\rr}$ thus leads to a stress tensor that
depends on the position of the subregion. As is usually the case for the
transition to a continuous description, in the following we assume that the
subregion is sufficiently big to perform the average but small with respect to
macroscopic lengths so that $\stress(\x)$ is a differentiable function.

To be more specific, in $d=3$ we consider two regions with different densities
that are separated by an interface. We orient the coordinate system so that
the interface normal points in $x$-direction. Translational invariance is
broken along the normal but still holds perpendicular so that $\stress(x)$ can
only depend on $x$. We consider a virtual transformation
\begin{equation}
  \vec h = \left(
    \begin{array}{ccc}
      \lam_\perp & 0 & 0 \\ 0 & \lam_\parallel & 0 \\ 0 & 0 & \lam_\parallel
    \end{array}\right)
\end{equation}
so that the box changes independently in the directions parallel and
perpendicular to the interface. The excess work thus becomes
\begin{equation}
  \delta\hat w_\text{ex} = \sum_R V_R \left[ \hat\sig\rr_{xx}\dd\lam_\perp +
    (\hat\sig\rr_{yy} + \hat\sig\rr_{zz})\dd\lam_\parallel \right],
\end{equation}
where we have split the total volume $V$ into non-overlapping stripes. The
changes of volume and interfacial area $A$ follow as
$\dd V_R = V_R\dd\lam_\perp + 2V_R\dd\lam_\parallel$ and
$\dd A = 2A\dd\lam_\parallel$, respectively, with $V_R=A\Delta x_R$ and
$\Delta x_R$ the width of stripe $R$. For fixed volume ($\dd V_R=0$) we thus
obtain $\delta\hat w_\text{ex}=\hat\gam\dd A$, whereby
\begin{equation}
  \label{eq:gam}
  \hat\gam(\z) \equiv \sum_R \Delta x_R \left[ -\hat\sig\rr_{xx} +
    \frac{1}{2} (\hat\sig\rr_{yy} + \hat\sig\rr_{zz}) \right]
\end{equation}
can be identified as an interfacial tension. Performing the average together
with the limit $\Delta x_R\ra 0$ of infinitesimal thin stripes thus leads to
the integral
\begin{equation}
  \label{eq:kb}
  \gam = \mean{\hat\gam} = \Int{x} [-\sig_\perp(x)+\sig_\parallel(x)]
\end{equation}
with normal stress $\sig_\perp(x)=\mean{\hat\sig^{(R)}}$ and tangential stress
$\sig_\parallel(x)=\mean{\hat\sig_{yy}^{(R)}}=\mean{\hat\sig_{zz}^{(R)}}$. In
bulk both stresses are equal and the integral only picks up contributions from
the interfacial region. We have thus derived another conjugate quantity, the
interfacial tension $\hat\gam(\z)$, which describes the work that would be
required to change the interfacial area while holding the total volume
constant. Eq.~(\ref{eq:kb}) is the same as what Kirkwood and Buff have derived
using mechanical arguments~\cite{kirk49}, which thus also holds in
non-equilibrium steady states.


\section{Effective swim forces}

We now apply the ideas developed so far to self-propelled particles. Every
particle is described by its position $\x_i$ and an orientation $\vec e_i$
with unity magnitude, $|\vec e_i|=1$, so that the full configuration is
specified by $\z\equiv\{\x_i,\vec e_i\}$. We adopt the following perspective:
by some mechanism available free energy (either stored or due to a local
gradient) is converted into motion; however, we do not resolve the details of
this mechanism explicitly. Rather, we exploit that due to this motion the
particles exert a force $\vec f_k(\z)$ onto the surrounding solvent, the
hydrodynamic drag, which in general depends on the details of the generated
flow profile. The role of the non-conservative forces $\vec f_k$ is thus to
describe the dissipation that is required to maintain the directed motion.

This perspective follows the approach of stochastic thermodynamics: we split
the total system into the actual system (the particles) and the environment
(the solvent). Particles are supplied with energy from the outside by some
means. For bacteria, one could image that some of that supplied ``fuel'' is
used for the internal metabolism, in which case the internal state has to be
included in the energy balance. On the other hand, colloidal particles are
inert and so all of this energy is dissipated into the solvent. To employ the
first law we need to model this dissipation. To this end we ask for the
effective force that would be necessary to achieve a certain speed. We stress
that this force is an effective force that does not necessarily appear in the
equations of motion (but see also Refs.~\citenum{hage15,yan15}).


\subsection{Active Brownian particles}

We now consider a more specific, minimal model for active colloids: active
Brownian particles (ABPs), which combines volume exclusion with persistence of
motion. This model neglects hydrodynamic interactions as well as long-ranged
phoretic interactions. The potential energy $U(\z)=\sum_{i<j}u(|\x_i-\x_j|)$
with conservative forces $\vec F_k=-\nabla_kU$ stems from pairwise
interactions with pair potential $u(r)$, and depends on particle positions
only. The equations of motion for the $N$ spherical particles read
\begin{equation}
  \label{eq:abp}
  \dot\x_k = v_0\vec e_k + \mu_0\vec F_k + \nois_k,
\end{equation}
where the noise $\nois_k$ models the solvent at temperature $T$ with
correlations
$\mean{\xi_{k,i}(t)\xi_{l,j}(t')}=2\mu_0\kT\delta_{kl}\delta_{ij}\delta(t-t')$.
Every particle is propelled with the same speed $v_0$ along its orientation
$\vec e_i$, which undergoes free diffusion with diffusion coefficient
$\Dr$. Since we neglect hydrodynamic interactions, the hydrodynamic forces are
simply
\begin{equation}
  \label{eq:abp:f}
  \vec f_k = -\frac{v_0}{\mu_0}\vec e_k \equiv -f_0\vec e_k,
\end{equation}
which is the frictional force due to a sphere moving with constant velocity
$v_0$.

The average rate for the housekeeping work Eq.~(\ref{eq:work:hk}) becomes
\begin{equation}
  \dot w_\text{hk} = -f_0\sum_{k=1}^N \mean{\vec e_k\cdot\dot\x_k} = -Nf_0v.
\end{equation}
To calculate the effective speed $v\equiv\mean{\vec e_k\cdot\dot\x_k}$ we
insert Eq.~(\ref{eq:abp}). The term $\mean{\vec e_k\cdot\nois_k}=0$ vanishes
since orientations and translational noises are uncorrelated. The other two
terms lead to $v=v_0-\mu_0\bar\rho\zeta$ with force imbalance coefficient
$\zeta$, see Ref.~\citenum{spec15} for details. The effective speed $v<v_0$ is
reduced compared to the free propulsion speed $v_0$ due to interactions with
other particles blocking the directed motion. However, on average particles
still move in the direction of the propulsion and hence $v>0$, which implies
that $\dot w_\text{hk}<0$ is negative. This exemplifies our argument that the
particles spent work on the solvent through the hydrodynamic drag.

\subsection{Pressure}

We now calculate the microscopic expression for the pressure of ABPs by
inserting the effective forces Eq.~(\ref{eq:abp:f}) into the stress
Eq.~(\ref{eq:stress}). For a uniform box change we thus obtain
\begin{equation}
  \label{eq:abp:p}
  \hat p(\z) = \bar\rho\kT + \hat p\si(\z) + \hat p\sa(\z),
\end{equation}
which can be written as sum of three contributions. The first term is the
ideal gas pressure with global density $\bar\rho\equiv N/V$. The second term
\begin{equation}
  \hat p\si(\z) = \frac{1}{2A}\sum_{k<l}^N w(|\x_k-\x_l|)
\end{equation}
with $w(r)\equiv-ru'(r)$ takes into account the direct interactions between
particles through the pair potential $u(r)$. It reduces to an expression that
manifestly involves only particle distances. The third term is the
contribution
\begin{equation}
  \hat p\sa(\z) = \frac{1}{2A}\frac{v_0}{\mu_0}\sum_{k=1}^N \vec e_k\cdot\x_k
\end{equation}
of the non-conservative forces Eq.~(\ref{eq:abp:f}) due to the
self-propulsion. Note that here the absolute particle positions enter. Our
approach via the excess work for a virtual deformation thus leads to the same
expression Eq.~(\ref{eq:abp:p}) for the pressure derived previously following
quite different arguments~\cite{taka14,solo15,solo15a,wink15}.


\section{Interfacial tension}
\label{sec:interface}

For sufficiently high densities and propulsion speeds, suspensions of ABPs
interacting via a repulsive pair potential separate into a dilute and a dense
phase, a phenomenon that strongly resembles passive liquid-gas phase
separation but has a purely dynamical
origin~\cite{yaou12,cate13,bial13,spec15}. In Ref.~\citenum{bial15} it has
been found that $\gam<0$ becomes negative, which implies that extending the
interface releases work. In a passive system this would lead to a
proliferation of interfaces and finally to homogenization, quite in contrast
to the observed stable phase separation of ABPs. This result is instructive
because it highlights two issues: (i)~not all properties that hold in
equilibrium are transferable to driven systems and (ii)~the role of
fluctuations. In equilibrium, the latter are determined by the free energy,
the same quantity that determines the pressure, cf. Eq.~(\ref{eq:eq:p}). This
connection no longer holds away from equilibrium.

\subsection{Constructing the interface}

\begin{figure}[t]
  \centering
  \includegraphics{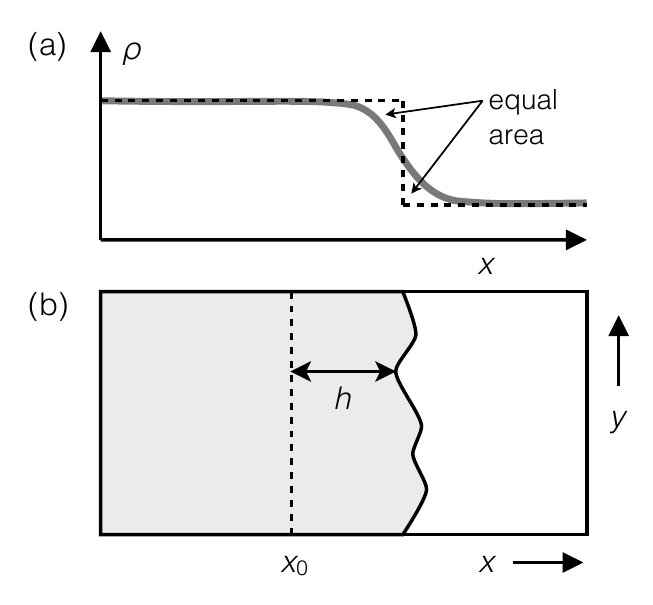}
  \caption{Gibbs dividing line in two dimensions. (a)~Average density profile
    (solid line) as a function of $x$. The position of the step profile
    (dashed line) is determined by an equal area rule. (b)~Instantaneous
    interface $x_0+h(y)$.}
  \label{fig:iface}
\end{figure}

Following Ref.~\citenum{bial15}, we consider a two-dimensional, large but
finite system with a single interface that spans the full system. This
interface is not static but changes constantly due to fluctuations. A
convenient concept is the Gibbs dividing surface constructed by extended the
bulk densities $\rho_\pm$ such that the position of the dividing line is
determined by the conservation of density, see
Fig.~\ref{fig:iface}(a). Without loss of generality, we assume that the normal
of the (averaged) interface coincides with the $x$-axis. We further assume
that we can construct the Gibbs dividing line for thin vertical stripes in
order to obtain an instantaneous interface $x=x_0+h(y)$ with respect to a
reference position $x_0$, see Fig.~\ref{fig:iface}(b). Note that we ignore
overhangs. Following capillary wave theory, in a finite system with dimensions
$L_x\times L_y$ and periodic boundaries, we decompose the interfacial profile
into complex Fourier modes
\begin{equation}
  h(y) = \sum_q h_q e^{\im qy}, \qquad h_q = \frac{1}{L_y}\IInt{y}{0}{L_y}
  h(y)e^{-\im qy}
\end{equation}
with $h_{-q}=h_q^\ast$, where $h_q^\ast$ denotes the conjugate complex. The
instantaneous length of the interface can then be calculated as
\begin{equation}
  \label{eq:ell}
  \hat\ell = \IInt{y}{0}{L_y} \sqrt{1+[\partial_yh]^2} \approx L_y +
  \frac{1}{2}L_y\sum_q q^2|h_q|^2,
\end{equation}
where we have approximated $\sqrt{1+x}\approx1+x/2$.

\subsection{Interfacial width}

In thermal equilibrium, the probability to observe an interface with length
$\ell$ is given by the change of free energy (\emph{viz.}, the reversible work
to deform the flat interface),
\begin{equation}
  \label{eq:p:ell}
  p(\ell) \propto e^{-[F(\ell)-F(L_y)]/\kT} \propto e^{-\gam_\text{eq}\ell/\kT}.
\end{equation}
For the second result we have expanded the free energy to linear order and
used $\gam_\text{eq}=\partial_\ell F|_{L_y}$,
cf. Eq.~(\ref{eq:eq:gam}). Following the usual argument, higher orders become
negligible in the limit of large system size since
$\partial_\ell^2F\sim\mathcal O(L_y^{-1})$.

In agreement with the observed stable phase separation, for ABPs we assume a
distribution $p(\ell)$ of interfacial lengths to exist. Following the same
argument as above, we can thus approximate
\begin{equation}
  p(\ell) \asymp e^{-\Om(\ell)} \propto e^{-\kap\ell}
\end{equation}
to leading order with unknown interfacial stiffness
$\kap=\partial_\ell\Om|_{L_y}>0$. Our use of the term ``stiffness'' is based
on its role in capillary wave theory for interface fluctuation in equilibrium,
where $\kap_\text{eq}=\gam_\text{eq}/\kT$ holds.

With this approximation we can determine the probability of the Fourier
coefficients $h_q$ in the non-equilibrium steady state. Employing
Eq.~(\ref{eq:ell}), to lowest order the modes are independent and Gaussian,
\begin{equation}
  \label{eq:h:P}
  P(\{h_q\}) \propto e^{-\kap\hat\ell}
  \propto \exp\left\{ -\frac{1}{2}\kap L_y \sum_q q^2 |h_q|^2 \right\},
\end{equation}
so that we can immediately read off the fluctuations
$\mean{|h_q|^2}=(\kap L_y q^2)^{-1}$. With this result we can finally
calculate the average width of the interface as
\begin{equation}
  w^2 = \frac{1}{L_y}\IInt{y}{0}{L_y} \mean{[h(y)]^2} = \sum_q \mean{|h_q|^2}.
\end{equation}
To leading order, the total width is then given by an unknown ``intrinsic''
width $w_0$ and the contributions due to the capillary waves with $|q|>0$ and
$q=(2\pi/L_y)n$,
\begin{equation}
  \label{eq:w}
  w^2 = w_0^2 + \frac{2}{\kap L_y}\sum_{q>0}\frac{1}{q^2} = w_0^2 +
  \frac{L_y}{12\kap},
\end{equation}
for which the linear dependence on $L_y$ has been confirmed
numerically~\cite{bial15}.


\subsection{Discussion}

To relate tension and stiffness, we start by rearranging their relation in
equilibrium,
\begin{equation}
  \label{eq:fdt:eq}
  \gam_\text{eq} = \frac{w}{\Delta\ell} = \kap_\text{eq}\kT,
\end{equation}
where $w$ is the (reversible) work required to extend the interface by
$\Delta\ell$. Note that this equation has the form of a
fluctuation-dissipation relation, where the tension (dissipation) is related
to the stiffness (fluctuations) through the thermal energy $\kT$. The physical
picture is that, for a fluctuation away from the flat profile, the system
``borrows'' energy from the heat reservoir through a collective fluctuation.

The conjecture of Ref.~\citenum{bial15} has been that a relation analogous to
Eq.~(\ref{eq:fdt:eq}) also holds for ABPs,
\begin{equation}
  \label{eq:fdt}
  \gam = \frac{w_\text{ex}}{\Delta\ell} = \kap\frac{\dot w_\text{hk}\tau}{N}
  = \kap(-f_0\lp),
\end{equation}
where the thermal energy is now replaced by the housekeeping work per particle
(during the time $\tau$ required to extend the interface by $\Delta\ell$),
which is given by the hydrodynamic friction force $f_0$ times the distance
$\lp(\tau)=v\tau$. The physical picture is that the housekeeping work sets the
scale for the energy that is available and thus determines the
fluctuations. Since $\kap>0$, Eq.~(\ref{eq:fdt}) indeed predicts a negative
tension $\gam<0$. In Ref.~\citenum{bial15} it has been shown that using
$\lp\approx v_0/\Dr$ leads to a quantitative agreement between the stiffness
$\kap$ extracted from measured interfacial widths using Eq.~(\ref{eq:w}) and
the tension $\gam$ calculated via Eq.~(\ref{eq:kb}).

\begin{figure}[t]
  \centering
  \includegraphics{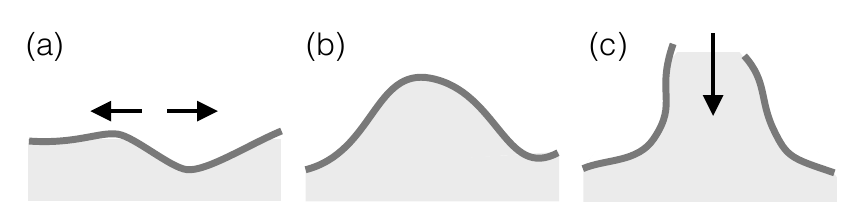}
  \caption{Fluctuations of the interface. (a)~Forces within the interface are
    extensile (arrows), for which fluctuations lead to (b)~buckling and
    finally (c)~rupture of the interface with particles entering (arrow).}
  \label{fig:buck}
\end{figure}

Fig.~\ref{fig:buck} sketches a possible scenario rationalizing a negative
tension with a positive stiffness based on the strong fluctuations of the
interface, which in simulations is seen to ``rupture'' and then to
reform. Since the tension is negative, the interface is extensile,
\emph{i.e.}, the active forces stretch the interface
[Fig.~\ref{fig:buck}(a)]. Fluctuations perpendicular to the interface
[Fig.~\ref{fig:buck}(b)] are thus amplified and finally the interface
``ruptures'' [Fig.~\ref{fig:buck}(c)]. Due to the polarization of particles at
the interface pointing towards to denser region, this is followed by an
incursion of particles stabilizing phase separation. Note that a somewhat
related scenario based on defects has been described for active
nematic-isotropic interfaces~\cite{blow14}. Moreover, simulations in three
dimensions, for which fluctuations are even stronger, show clear evidence for
these collective incursions from the interface into the dense
domains~\cite{wyso14}.


\section{Conclusions}

In this letter we have introduced two fundamental concepts: (i)~conjugate
observables out of equilibrium based on (fluctuating) work, and (ii)~the use
of effective non-conservative forces to model the dissipated work necessary to
sustain the directed motion of active matter. As a first step we have applied
these ideas to the minimal model of active Brownian particles. The result for
the pressure agrees with previous derivations. For the interfacial tension we
have proposed a relation connecting fluctuations with the housekeeping
work. While here we have considered large systems neglecting boundaries and
walls, these have a profound influence on active matter and will have to be
addressed in future work.


\acknowledgments

The DFG is acknowledged for financial support within priority program SPP 1726
(grant number SP 1382/3-1).


\end{document}